\documentclass[prl,twocolumn,preprintnumbers ,amsmath, amssymb, superscriptaddress, aps]{revtex4-1}

\usepackage{color}
\usepackage{amsmath}
\usepackage{pifont}
\usepackage{amssymb}  
\usepackage{bbold}
\usepackage{float}
\usepackage{tikz}
\usepackage{makecell}
\usepackage{subfigure}
\usepackage{pifont}   
\usepackage{graphicx} 
\usepackage{dcolumn}  
\usepackage{bm}       
\usepackage{multirow} 
\usepackage{hyperref}
\usepackage{placeins}
\newcommand{\vect}[1]{\mathbf{#1}}

\begin{document}

\renewcommand\floatpagefraction{0.8} 
\renewcommand\topfraction{0.8}       

\author{Michael Vogl}
\affiliation{Department of Physics, The University of Texas at Austin, Austin, TX 78712, USA}
\author{Martin Rodriguez-Vega}
\affiliation{Department of Physics, The University of Texas at Austin, Austin, TX 78712, USA}
\affiliation{Department of Physics, Northeastern University, Boston, MA 02115, USA}
\author{Gregory A. Fiete}
\affiliation{Department of Physics, Northeastern University, Boston, MA 02115, USA}
\affiliation{Department of Physics, Massachusetts Institute of Technology, Cambridge, MA 02139, USA}
\title{Floquet engineering of interlayer couplings: \\Tuning the magic angle of twisted bilayer graphene at the exit of a waveguide}
\date{\today}

\begin{abstract}
We introduce a new approach that allows one complete control over the modulation of the effective twist angle change in few-layer van der Waals heterostructures by irradiating them with longitudinal waves of light at the end of a waveguide.  As a specific application, we consider twisted bilayer graphene and show that one can tune the magic angles to be either larger or smaller, allowing in-situ experimental control of the phase diagram of this and other related materials.  A waveguide allows one to circumvent the free-space constraints on the absence of longitudinal electric field components of light.   We propose to place twisted bilayer graphene at a specific location at the exit of a waveguide, such that it is subjected to purely longitudinal components of a transverse magnetic modes (TM) wave.
\end{abstract}
\maketitle

\textit{Introduction.} Van der Waals heterostructures are flexible platforms to engineer systems with designer properties because they are not constrained by the chemistry of the two-dimensional materials composing the system \cite{Geim2013, Novoselovaac9439}. The properties of such heterostructures can be tuned with the knobs provided by the layer material choice, stacking order, temperature, pressure, applied fields, time-dependent drives, and other means. Recent experiential advances made possible the introduction of a new knob: a relative twist angle between the layers. In particular,  twisted bilayer graphene (TBG) has emerged as a controllable platform for the study of strongly correlated physics\cite{Wu_2018,Cao2018sc,Codecidoeaaw9770,Yankowitz1059,chichinadze2019nematic,Chou_2019,Guinea13174,PhysRevLett.122.257002,PhysRevB.99.134515,caldern2019correlated,saito2019decoupling,stepanov2019interplay,Kang_2019,Volovik_2018,Po_2018,Ochi_2018,Gonz_lez_2019,Sherkunov_2018,Laksono_2018,Venderbos_2018,PhysRevB.81.165105,Seo_2019}. The interplay of the interlayer tunneling amplitude, and the induced moir\'e pattern lead to the formation of isolated flat bands at specific \textit{magic} angles, where interaction effects can dominate \cite{Bistritzer12233,PhysRevB.93.035452}. Specifically, TBG exhibits an insulating state at half-filling driven by electron-electron interactions \cite{Cao2018,Kim3364}, which can transition into superconducting~\cite{Cao2018sc}, and ferromagnetic~\cite{Sharpe605,Seo_2019} states at specific filling factors. 

One of the main parameters that control the location of the magic angle is the interlayer tunneling amplitude. Fine-tuning of the rotation angle represents one of the main challenges for the experimental realization of strongly correlated states, since the band structure is very sensitive to small changes in this angle. Recent theoretical studies suggested the possibility to overcome small misalignments away from the magic angle by applying uniaxial pressure \cite{PhysRevB.98.085144, Chittari_2018}. Furthermore, experiments have already shown that hydrostatic pressure can increase the tunneling amplitude between the layers and lead to strongly correlated states in samples otherwise dominated by the kinetic energy \cite{Yankowitz2018, Yankowitz1059}. This procedure addresses the layer misalignment issue for samples with angles larger than the magic angle. However, no controlled scheme has been proposed to date to reduce the tunneling amplitude between the layers and decrease the magic angle. 

\begin{figure}[t]
	\centering
	\includegraphics[width=.85\linewidth]{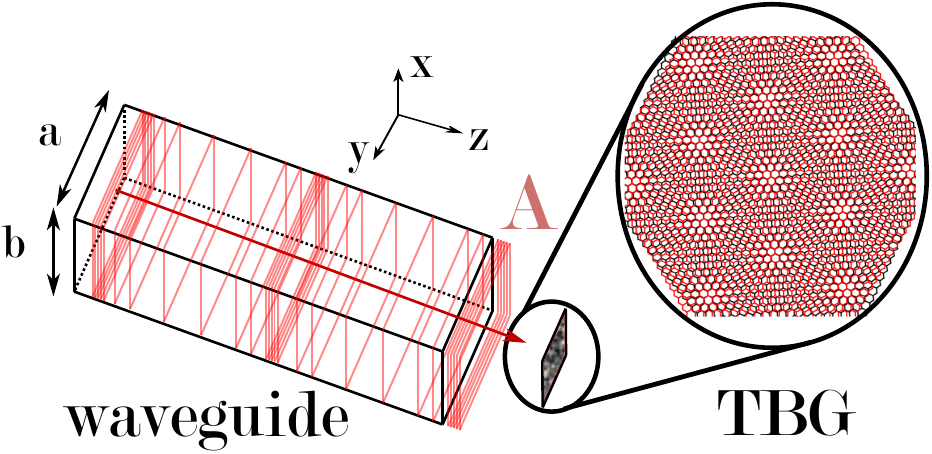}
	\caption{Cartoon of twisted bilayer graphene placed at the exit of a rectangular waveguide. Here, $a$ and $b$ denote the dimensions of the waveguide and $\vect A$ the vector potential inside the waveguide. The red square pattern highlights its longitudinal nature.}
	\label{fig:placeholdertwisted}
\end{figure}

Traditional Floquet engineering in free space using circularly-polarized light like in the case of graphene\cite{PhysRevB.79.081406,PhysRevB.95.125401,PhysRevB.93.115420,kristinsson2016control,PhysRevB.81.165433,PhysRevB.89.121401,PhysRevB.90.115423,PhysRevA.91.043625,mciver2020light} allows one to break time-reversal symmetry, induce non-trivial gaps in the quasienergy spectrum and leads to rich topological phase diagrams by modifying the intralayer hopping amplitude~\cite{PhysRevB.79.081406,Lindner2011}. Recently, this traditional Floquet protocol was applied to twisted bilayer graphene. It allows one to induce topological transitions at large twist angles using high-frequency drives~\cite{PhysRevResearch.1.023031}, tune the bandwidth of the energy levels near charge neutrality and induce a transition to a Chern insulating phase with Chern number $C=4$~\cite{li2019floquetengineered} in the high-frequency regime, and induce flat bands using near-infrared light in a wide range of twist angles~\cite{katz2019floquet}. Despite the high degree of tunability induced by the Floquet protocols and related ones that make use of linearly polarized light \cite{Yudin_2016}, the interlayer tunneling amplitude is not easily affected. 

In this work, we demonstrate that a waveguide transverse magnetic (TM) mode can directly couple to the inter-layer tunneling in few-layer van der Waals systems and be used to tune the magic angle in-situ, allowing complete control over the effective rotation angle and the corresponding phase diagram.  Although this work focuses on twisted bilayer graphene, the drive protocol we propose is general and can be applied to other low-dimensional systems to tune the strength of interlayer couplings \cite{fleischmann2019moir}. Our approach increases the degree to which the parameters of low-dimensional systems can be modified in out-of-equilibrium settings. It allows one to directly tune the Fermi velocity of the quasienergy spectrum and it increases the degree to which the parameters of low-dimensional systems can be modified in a controlled way. 




\textit{Static model for bilayer graphene.} When two graphene layers are misaligned with respect to each other by a rotation, a moir\'e pattern emerges with a tunneling amplitude between the layers that is dominated by processes at lattice sites located directly on top of each other. The low-energy effective Hamiltonian for static twisted bilayer graphene (TBG) is \cite{Bistritzer12233,Wu_2018,Rost_2019,fleischmann2019perfect,fleischmann2019moir,xie2018nature,PhysRevResearch.1.013001}
%
\begin{equation}
H=\begin{pmatrix}
h(-\theta/2,\vect k-\kappa_-)&T(\vect x)\\
T^\dag(\vect x)&h(\theta/2,\vect k-\kappa_+)
\end{pmatrix},
\end{equation}
which describes two stacked graphene layers that are twisted with respect to each other by an angle $\theta$. Each graphene layer is described by the Hamiltonian
\begin{equation}
h(\theta,\vect k)=\beta\begin{pmatrix}
0&f(R(\theta)\vect k)\\
f^*(R(\theta)\vect k)&0
\end{pmatrix},
\label{TwistHam}
\end{equation}
where $\vect k$ is the in-plane momentum and $\beta$ sets the strength of the intralayer hopping.  The function $R(\theta)$ describes a rotation around the axis perpendicular to the graphene sheets by an angle $\theta$, and $f(\vect k)=e^{-2 i a_0 k_y/3}+2 e^{ i a_0 k_y/3} \sin \left( a_0 k_x/\sqrt{3}- \pi/6\right)$ is the intra-layer hopping matrix expressed in momentum space with $a_0$ the lattice constant of graphene. In each layer, we shift the momenta by the $K$-points $\kappa_\pm=k_\theta\left(-\sqrt{3}/2,\pm 1/2\right)$. Here $k_\theta=k_D \sin\left(\theta/2\right)$ and $k_D=8\pi/(3a_0)$ defines the angle dependent momentum scale of the bilayer system. The tunneling processes between the layers are approximated by
$
T(\vect x)=\sum_{i=-1}^1 e^{-i\vect b_i\vect x} T_i,
$
where the matrix structure is given by
\begin{equation}
T_n=w_0\mathbb{1}_2+w_1\left(\cos\left(\frac{2\pi n}{3}\right)\sigma_1+\sin\left(\frac{2\pi n}{3}\right)\sigma_2\right),
\end{equation}
where $ \vect b_{\pm 1}= k_\theta\left(\pm \sqrt{3},3\right)/2$ are the reciprocal moir\'e lattice vectors and $\vect b_0=(0,0)$ is introduced to write $T(\vect x)$ compactly. We stress that the $w_i$ are proportional to interlayer hopping elements in a corresponding tight binding model \cite{Rost_2019}.

In bilayer graphene, some stacking configurations are energetically more favorable over others. Particularly, AB/BA stacking, where half the atoms of the upper graphene layer lie on top of the empty hexagon centers  of the lower layer, is favored over AA stacked regions where the upper layer is directly on top of the lower layer. This energy-dependent stacking configuration leads to relaxation effects in twisted bilayer graphene, where AB/BA regions are favored over AA regions~\cite{PhysRevB.96.075311,fleischmann2019moir}.  We take this effect and corrugation into account through the additional parameter $w_1$ in the tunneling amplitude \cite{katz2019floquet,li2019floquetengineered}. This can be understood by recognizing that the parameter $w_0$ enters the Hamiltonian in the form of a AA stacking hopping. On the other hand, $w_1$ enters as hopping in bilayer graphene AB/BA configuration. Therefore, $w_1/w_0>1$  corresponds to systems where $AB$-stacked regions are favored. Then, the ratio $w_1/w_0$ can be used to model systems with larger or smaller AB/BA regions~ \cite{katz2019floquet,li2019floquetengineered}. In appendix \ref{app:numeric}, we describe the numerical band structure implementation. Throughout this work, we use the parameters 
$\beta=\hbar v_F/a_0=2.36$ eV, $a_0 = 2.46\mbox{ \normalfont\AA}$ and $w_1=110$ meV. 

\begin{figure}[t]
	\centering
	\includegraphics[width=7.5cm]{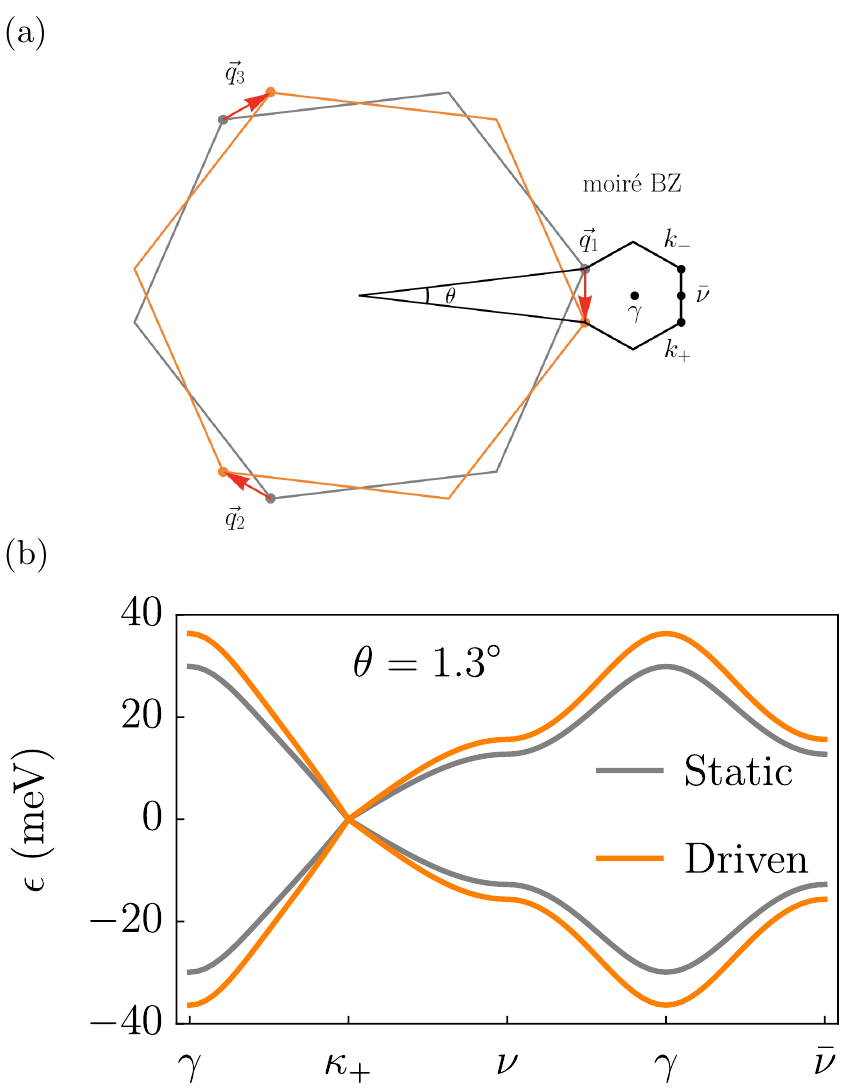}
	\caption{(a) Twisted bilayer graphene moir\'e Brillouin zone (mBZ). Large orange and black hexagons represent the graphene layer BZs. (b) TBG quasienergy spectrum in a waveguide along a high-symmetry path in the mBZ (orange). The gray curve corresponds to the static case. The specific parameters are $w_0=0$, $\theta = 1.3^\circ$, $\Omega/W=1.3$, and $a_{AB} A= 0.4$. We estimated the bandwidth as $W\approx \beta$. The plot was made by taking into account three Floquet copies, which was enough for convergence.}
	\label{fig:bands}
\end{figure}

\textit{Driving protocol.}
We now place the twisted bilayer graphene sample at the exit of a metallic waveguide, as sketched in Fig. \ref{fig:placeholdertwisted}. The transverse magnetic modes (TM) are derived from the vector potential $\vect A = \hat z A \sin\left(m\pi x/a\right)\sin\left( n \pi y / b \right) \mathrm{Re}(e^{-ik_z z-i\Omega t})$, where $k_z=\sqrt{k^2-(m\pi/a)^2-(n \pi/ b)^2}$ is the wavenumber in the $z$-direction, $k^2=\Omega^2 \mu \varepsilon$, $\mu$ is the permeability constant of the insulator inside the waveguide, $m,n\in \mathbb{Z}$ characterize the transverse modes and $\varepsilon$ is the dielectric constant. The waveguide cut-off frequency is given by $\Omega_c = 1/(\mu \varepsilon) \sqrt{(m\pi/a)^2+(n \pi/ b)^2} $. A full derivation and the full electric and magnetic fields inside the waveguide are given in the appendix \ref{transverse_fields}.

The maxima of the sine functions are ideal locations to place the TBG sheet because the field has a maximum amplitude $A$ at these locations and because the vector potential is constant to second order in the $x,y$ direction. We stress that only the properties of $\vect A$ (as opposed to the physical fields $\vect E$ and $\vect B$) matter because this is what enters the Hamiltonian. Taking $\vect A=A \mathrm{Re}(e^{-ik_z z-i\Omega t}) \hat z$ therefore is a good approximation if the sample is small compared to the waveguide dimensions $a,b$. 

At the tight binding level, $\vect A$ enters the Hamiltonian through the hopping elements via the Peierls substitution $t_{ij}\to t_{ij} \exp\left(-i \int_{\vect r_i}^{\vect r_j} \vect A \cdot d\vect l \right)$, where the line integral is associated with the interlayer hopping direction in real space and $\vect r_i$ labels site $i$. The continuum model we discuss here is an approximation to a tight binding model and the $w_i$ that appear in the continuum model correspond to the interlayer couplings of the underlying tight binding description. We may therefore introduce the vector potential via the $w_i$. The vector potential we consider only has a component in $z$- direction and therefore only influences interlayer hopping amplitudes. For computational simplicity, we will assume that the exit of the cavity is at $z=0$ and that the twisted bilayer graphene sheet is placed at such a location. We note that the coupling of the longitudinal vector potential is not specific to twisted bilayer graphene and applies to other van der Waals heterostructures.

In an idealized scenario, the additional phase factor in the tunneling amplitude is uniform across the sample. However, lattice relaxation effects and surface roughness introduce a position dependence to the phase due to different layer separations. An exact treatment of such a complicated scenario is not the purpose of this work. In order to simplify the picture and make analytic progress, let us recall that relaxation effects lead to two different hopping amplitudes: $w_1$, the hopping on $AB$ stacking and $w_0$, the hopping of $AA$ stacking. Therefore, to leading order, the two hopping amplitudes acquire the phases
\begin{align}
	w_1 & \to w_1e^{-i a_{AB} A\cos(\Omega t)},\\
	w_0 & \to w_0e^{-ia_{AA} A\cos(\Omega t)},
	\label{couplings}
\end{align}
where $a_{AA}=3.6\mbox{ \normalfont\AA}$ is the distance of the graphene layers in $AA$-stacked regions and $a_{AB}=3.4\mbox{ \normalfont\AA}$ the corresponding quantity for $AB$ stacking \cite{PhysRevB.99.205134}. For other types of low-dimensional materials, an analogous replacements is valid for their interlayer coupling elements.

Using an effective time-independent Floquet Hamiltonian to describe the system simplifies our discussion. To lowest order, the high frequency approximation of a Floquet Hamiltonian (in $1/\Omega$) is the time average over one period $2 \pi/\Omega$, $\hat H_F = \int_0^{2\pi/\Omega} \Omega ds/(2\pi) \hat H(s) $. This type of approximation can be expected to be good when $W< \Omega$\cite{Abanin_2017,Eckardt_2015,Mikami_2016}, where $W$ is the bandwidth of the Hamiltonian. In the present case, the only time dependence of $\hat H$ is through the hopping amplitudes in Eq. \eqref{couplings}. Furthermore, the hopping amplitudes $w_i$ for both the continuum model we discuss and the corresponding tight binding model enter linearly into the Hamiltonian i.e. $H=H_{C}+w_1H_1+w_0H_0$, where $H_C$ is the $w_{0,1}$- independent part of $H$. Therefore, the lowest order Floquet-Magnus approximation is found simply by taking the time average of $w_1$ and $w_0$. We find that the effective Floquet Hamiltonian has the same structure as the static Hamiltonian Eq. \eqref{TwistHam}, with the hopping amplitudes renormalized as 
\begin{align}
	w_1 \to & \tilde w_1=J_0\left(\left| a_{AB} A\right|\right) w_1,\\
	w_0\to & \tilde w_0= J_0\left(\left| a_{AA} A\right|\right)w_0,
\end{align}
where $J_0$ is the zeroth Bessel function of the first kind. This light-induced direct renormalization of the interlayer hopping amplitude is one of the main result of our work. We stress that the result is only valid in the large frequency limit; we turn to the low frequency regime below. The same driving protocol can be used to selectively weaken the inter-layer couplings of other few-layer materials such as multilayer graphenes, transition metal dichalcogenide multilayers, phosphorene multilayers and others that have been of recent interest \cite{Cheng_2019,Liu_2014,wu2019anomalous,PhysRevB.96.035442,shang2019artificial,Abdullah_2017,Iorsh_2017}.

In Fig. \ref{fig:bands}, we plot the quasienergies along a high-symmetry path in the mBZ. The quasi-energies in the driven case were computed using an extended space Floquet Hamiltonian \cite{Eckardt_2015} and they agree well (no disagreement on the scale of the plot) with our effective Hamiltonian. We find that the effect of the drive on the low quasi-energy band structure is an increased Fermi velocity near $\kappa_+$. Unlike the case of Floquet drives in free space with circularly polarized light, our protocol does not open gaps in the quasienergy spectrum because time-reversal symmetry is not broken.  

\begin{figure}[t]
	\centering
	\includegraphics[width=7.0cm]{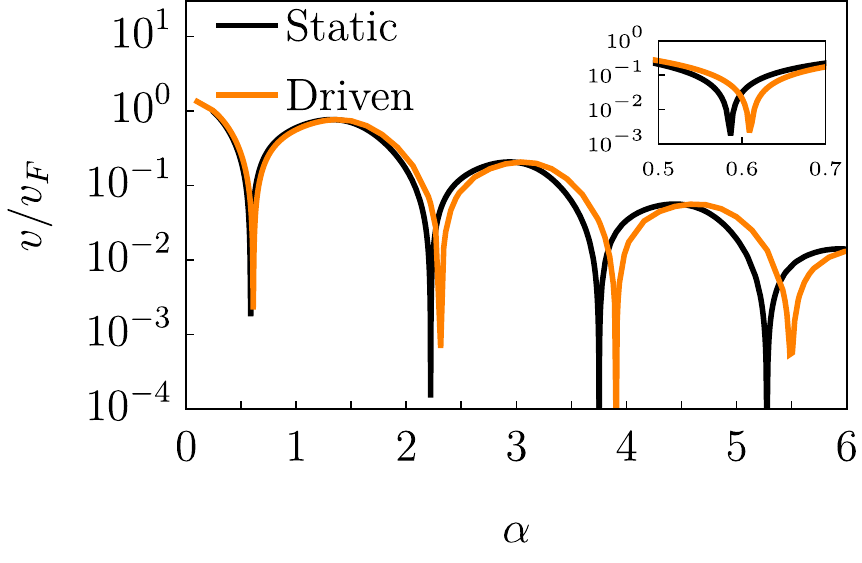}
	\caption{Renormalized effective Floquet Fermi velocity near $\kappa_+$ of the quasienergy spectrum as a function of the inverse twist angle between the layers, $\alpha=w_1/(2v_Fk_D\sin(\theta/2))$ (orange). The static case is shown in black for reference. The drive parameters are  $a_{AB} A=0.4$ and $\Omega/W=1.3$. Here we estimated the half bandwidth as $W\approx \beta$. The plot was made by taking into account three Floquet copies, which was enough for convergence.}
	\label{fig:twistangle}
\end{figure}

Now we discuss the effect of the TM modes on the magic angles. To simplify the discussion, we consider the chiral case, where $w_{0}=0$. Furthermore, we assume small energies such that the Hamiltonian \eqref{TwistHam}  near one of the K points of graphene becomes linear in momenta i.e. $f(\vect k)\approx a_0(k_x-ik_y)$. In this case there is only one dimensionless parameter that enters the Schr\"odinger equation, $\alpha=w_1/(2v_Fk_D\sin(\theta/2))$. In Ref.~[\onlinecite{PhysRevLett.122.106405}], Tarnopolsky \textit{et. al.} considered this limit in equilibrium, and found perfectly flat bands appearing for $\alpha_1\approx0.586$ or $\theta_1\approx 1.09^\circ$. They found that further flat bands exist for $\alpha_n=\alpha_1+n\Delta\alpha$ with $\Delta \alpha = 3/2$ and $n\in \mathbb{N}$. With our modified values for $w_1$ we therefore find that the magic angles are to good approximation  $\sin(\theta/2)\approx\theta/2$ given as

\begin{equation}
	\theta_n=\frac{w_1J_0\left(\left| a_{AB} A\right|\right)}{v_F k_D \alpha_n}.
	\label{eq:magic_angle_driven}
\end{equation}

Therefore, the degree of tunability of the hopping amplitude depends on the range of values $\eta \equiv (e/\hbar) a_{AB} A$ can take in experiments, where we reintroduced $\hbar$ and $e$ to have $A$ given in more conventional units. In terms of the electric field amplitude, we have $\eta \equiv e a_{AB} E/(\hbar \Omega)$. Then, in a pump-probe setup, with a pump drive frequency $f=\Omega/(2\pi) = 650$~THz ($\sim2.7$~eV,  in the range of blue light), and electric field strength $E=15$MV/cm, we obtain $\eta \approx 0.19$, which leads to $\theta_F/\theta = 0.99 $. 

Stronger laser pulses with electric fields up to $E=25$ MV/cm lead to $\theta_F/\theta =0.975$. Recent experiments in graphene have employed peak electric fields of up to $\sim 30 MV/cm$ with frequencies in the near-IR regime ($\sim 375$ THz)~\cite{Higuchi2017,PhysRevLett.121.207401}. Furthermore, current laser technology can reach field strengths on the order of $10$ MV/cm in the UV regime~\cite{li2019floquetengineered}. The pump frequency is chosen to be larger than half the bandwidth. While this is smaller than the bandwidth, which is approximately the point where a high-frequency approximation for the center bands starts breaking down globally \cite{Eckardt_2015,Abanin_2017,Mikami_2016,Bukov_2015,PhysRevX.9.021037,PhysRevB.101.024303,Rodriguez_Vega_2018,Vogl_2019}, it is not important for us since we only are interested in center flat bands. These bands are accurately described for much lower frequencies until around half the bandwidth. Particularly we verified this numerically using an extended space picture method \cite{Eckardt_2015}. The predictions from our time-independent model agree perfectly with numerics. The value $\eta$ can be further tuned by changing the permitivity inside the cavity. A waveguide of size of the order of $a=1 $ mm can support frequencies as small as $1$~THz when the filling insulator is air. Therefore, the high-frequency regime necessary for the validity of Eq.~\ref{eq:magic_angle_driven} is allowed in a typical waveguide. 
 
In Fig. \ref{fig:twistangle}, we plot the Fermi-velocity of the lowest band $\epsilon_k^{(1)}$
\begin{equation}
	v=\sqrt{(\partial_{k_x}\epsilon_k^{{(1)}})^2+(\partial_{k_y}\epsilon^{(1)}_k)^2}|_{\kappa_+}
\end{equation} 
near $\kappa_+$ as a function of parameter $\alpha$. The quasi-energies were calculated using the extended space representation of the Floquet Hamiltonian \cite{Eckardt_2015}.

We find that the points where the Fermi velocity vanishes, and where we have flat bands is shifted to larger values of $\alpha$, or smaller angles $\theta$. In the inset, we show that the position of the flat bands for the first magic angle has shifted to $\tilde \alpha=0.608$. This shift  is a $4\%$ change in the magic angle, and in good agreement with our analytical estimates \eqref{eq:magic_angle_driven}. 

\begin{figure}[t]
	\centering
	\includegraphics[width=0.49\linewidth]{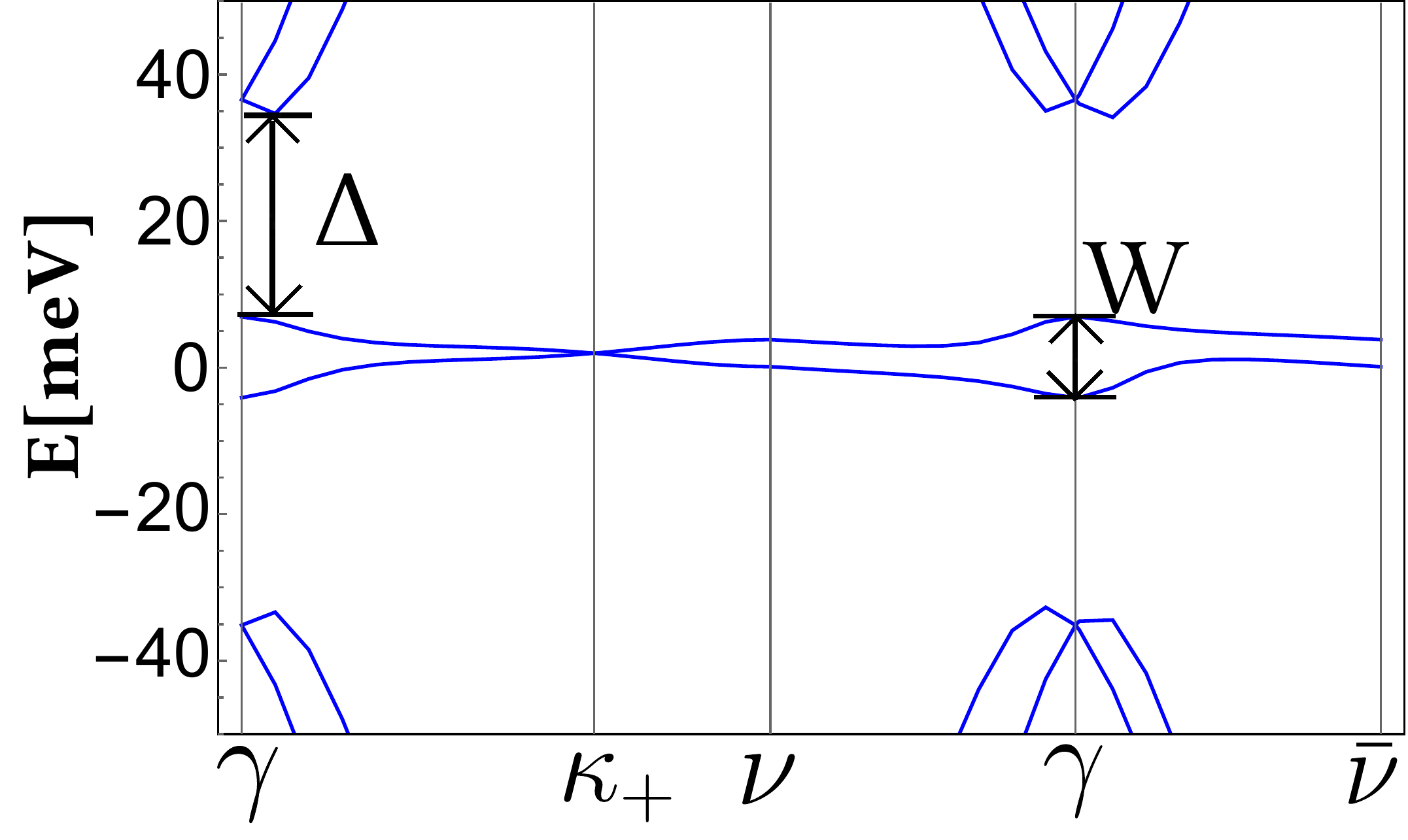}
	\includegraphics[width=0.49\linewidth]{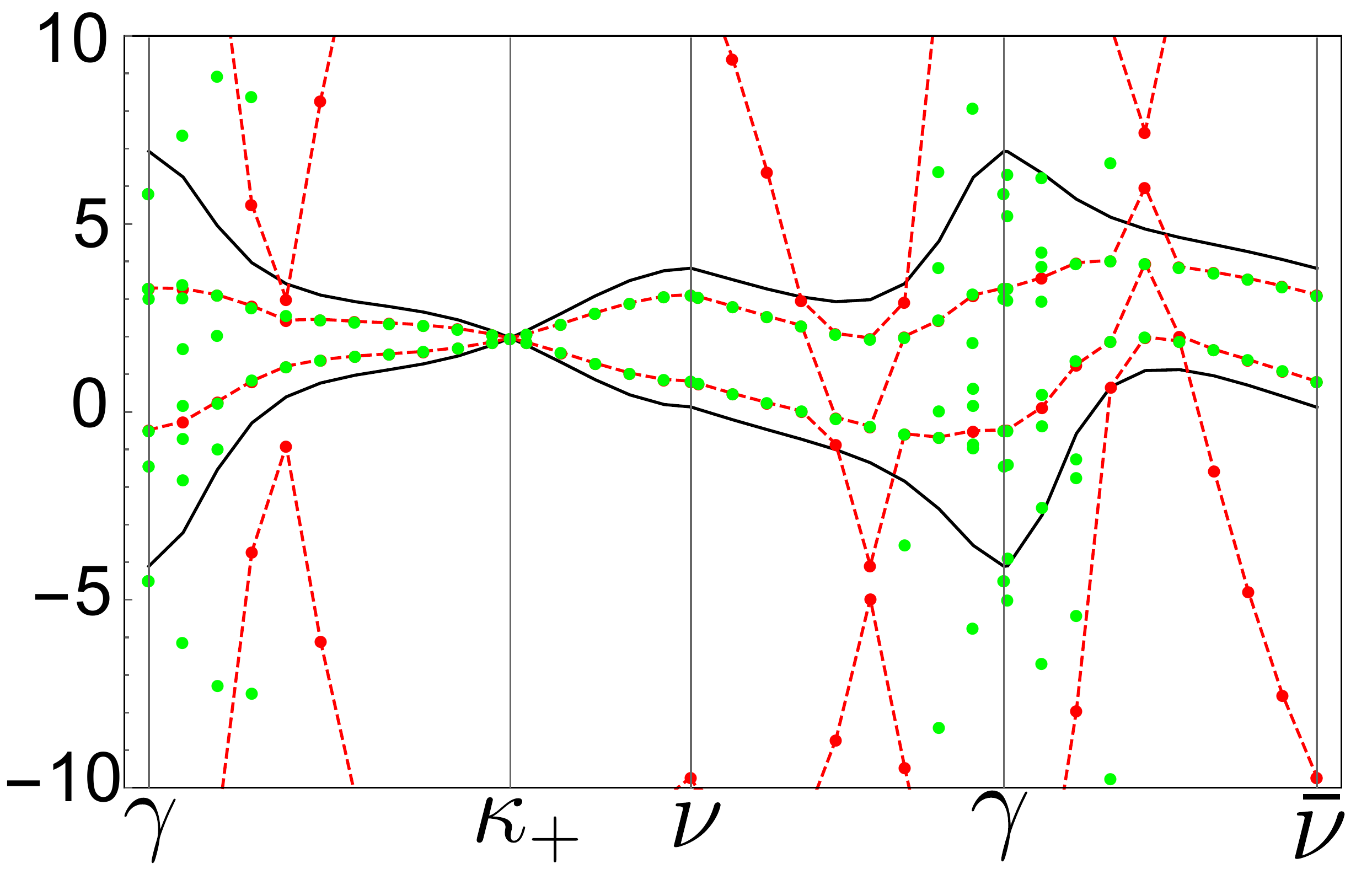}
	\caption{{\bf Left:} Band structure of twisted bilayer graphene at twist angle $1.15^\circ$, $w_0=88$ meV, and $w_1=110$ meV. {\bf Right:} The center Floquet bands. The system is driven by far infra red light with a frequency $\Omega=20$ meV  or $\sim 4.8$ THz and strength $Aa_{AA}=1.26$, which corresponds to an electric field $E\sim0.7$ MV/cm. Black curves, undriven system; Red 5 Floquet copies; Green 7 Floquet copies. The points where red and green points agree correspond to the same Floquet copy and are converged. Other stray points correspond to unconverged higher Floquet copies and are not important to the energy states around 0.}
	\label{fig:gapopening}
\end{figure}

\textit{Low frequency window. } Thus far we have only looked at the limit of large frequencies. However,  near the magic angle when $w_0<w_1$, there exists a gap that opens around the nearly flat bands. The band gap is such that driving frequencies that fulfill $W<\Omega<\Delta$, where $W$ is the bandwidth of the flat band and $\Delta$ the size of the band gap, are possible as shown in Fig. \ref{fig:gapopening}.  For such a regime it is possible to numerically find clear answers as to what happens to the lowest flat bands, as indicated in the right panel of Fig. \ref{fig:gapopening}.

From the converged lowest Floquet copy where red and green points coincide we can see that the band for angles above the magic angle of $1.10^\circ$ has been flattened - unlike what happens in high frequency regime, where it becomes steeper. This indicates that the flat bands are shifted to larger angles. Indeed we find numerically that flat bands appear at $1.12^\circ$ rather than $1.10^\circ$ as they normally would for $w_0=88$ meV and $w_1=110$ meV. Here, we used a relatively weak  electric field $E\sim0.7$ MV/cm. Strong fields of up to 100 MV$/$cm have been achieved in the literature in the range 15-50 THz \cite{kampfrath2013, sell2008}, which allows for a wide range of experimental tunability. Therefore, depending on driving frequencies we can either increase or lower the magic angle also for the realistic case of $w_0\neq 0$. 


\textit{Conclusions.} We introduced a new mechanism to dynamically tune the effective interlayer tunneling amplitude in few-layer quantum materials using longitudinal electric fields available in a waveguide. This allows one complete control over the modulation of the effective twist angle change in few-layer van der Waals heterostructures in-situ with intensities currently available in experiment.  

\begin{acknowledgments}
	We thank Fengcheng Wu for useful discussions and Sam Shallcross for useful comments on our manuscript. This work was primarily supported by the National Science Foundation through the Center for Dynamics and Control of Materials: an NSF MRSEC under Cooperative Agreement No. DMR-1720595.  Partial support also from NSF Grant No. DMR-1949701.
\end{acknowledgments}

\bibliographystyle{unsrtnat}
\bibliography{LongiTudLightTwist}

\appendix
\section{Numerical solution of the Hamiltonian}
\label{app:numeric}

In this appendix, we detailed the numerical implementation of the static Hamiltonian for twisted bilayer graphene. We employ the \textit{cellular method}, that was first introduced in 1933 by Wigner and Seitz \cite{PhysRev.43.804}. Below we give a short summary of the steps needed. 

Our starting point is the Schr\"odinger equation for a periodic Hamiltonian
\begin{equation}
E\psi=H(\hat p,x)\psi,
\end{equation}
where $\hat p=-i\nabla$. Since $H(p,x)$ is periodic, with the same periodicity of the lattice, it commutes with the translation operators $T_X=e^{i\hat p X}$ for lattice vectors $X$. That is, it also fulfills the eigenvalue equation
\begin{equation}
T_X\psi(x)=C\psi(x),
\end{equation} 
which has the eigenvalues $C=e^{ikX}$ and therefore one may rearrange the equation as $e^{-ikX}\psi(x+X)=\psi(x)$. If we multiply both sides by $e^{-ikx}$ we see that $u(x)=e^{ikx}\psi(x)$ is periodic, which is known a Bloch's theorem. Therefore, setting $\psi(x)=e^{ikx} u(x)$ in the Schr\"odinger equation gives us
\begin{equation}
E_k u(x)=H(p+k,x)u(x),
\end{equation}
which describes bands $E_k$ and has coordinates restricted to one lattice unit cell. The scalar product is restricted in a similar fashion. Calculating bands therefore merely amounts to writing the Hamiltonian in a plane wave basis
\begin{equation}
H_{\vect n \vect m}(k)=\frac{1}{V_{UZ}}\int_{\mathcal{A}_{UZ}} e^{i\vect Q_n \vect x}H(k,p,x) e^{-i\vect Q_m \vect x}d^nx,
\end{equation}
where $\mathcal{A}_{UZ}$ is a unit cell, $V_{UZ}$ the volume of the unit cell and $\vect Q_m$ are reciprocal lattice vectors. The integral over the $\mathcal{A}_{UZ}$ can be performed analytically, leading to efficient implementations of the band structure calculation.

\section{Electric and magnetic transverse fields.}
\label{transverse_fields}
The wave equation for the vector potential $\vect A$ in free space according to Maxwell's equations is given as
\begin{equation}
	\nabla^2 \vect A =\mu\epsilon \partial_t^2 \vect A 
\end{equation}

A  specific solution to this equation is chosen as
\begin{equation}
\vect A = \hat z A \sin\left(k_x x\right)\sin\left( k_y y\right) \mathrm{Re}(e^{-ik_z z-i\Omega t}),
\end{equation}
where $k_z=\sqrt{k^2-k_x^2-k_y^2}$ and $k=\Omega \sqrt{\mu\epsilon}$.
We recall that 
\begin{equation}
	\begin{aligned}
	\vect H=\nabla\times \vect A;\quad \partial_t \vect E=\frac{1}{\epsilon}\nabla\times \vect H
	\end{aligned}
\end{equation}
From here one may see that the solution for $\vect A$ we found is enough to accommodate the boundary conditions 
\begin{equation}
	E_y(x=a)=E_z(x=a)=E_x(y=b)=E_z(y=b)=0
\end{equation}
that a rectangular metallic waveguide introduces if we choose $k_x=\frac{m\pi}{a}$ and $k_y=\frac{n\pi}{b}$ with $n,m\in \mathbb{Z}$.
This solution is called a transverse magnetic mode (TM) of the cavity and the fields are given as
\begin{equation}
	\begin{aligned}
	&E_x\propto \cos\left(\frac{m\pi x}{a}\right)\sin\left(\frac{n\pi y}{b}\right)\mathrm{Re}(e^{-ik_z z-i\Omega t})\\
	&	E_y\propto \sin\left(\frac{m\pi x}{a}\right)\cos\left(\frac{n\pi y}{b}\right)\mathrm{Re}(e^{-ik_z z-i\Omega t})\\
	&E_z\propto\sin\left(\frac{m\pi x}{a}\right)\sin\left(\frac{n\pi y}{b}\right)\mathrm{Re}(e^{-ik_z z-i\Omega t})\\
	&H_x\propto \sin\left(\frac{m\pi x}{a}\right)\cos\left(\frac{n\pi y}{b}\right)\mathrm{Re}(e^{-ik_z z-i\Omega t})\\
	&H_y\propto \cos\left(\frac{m\pi x}{a}\right)\sin\left(\frac{n\pi y}{b}\right)\mathrm{Re}(e^{-ik_z z-i\Omega t})\\
	&H_z=0
	\end{aligned}.
\end{equation}

At the points we want to place the twisted bilayer graphene sheets we have $\sin=1$ and $\cos=0$ and therefore only $E_z$ is non-zero. This coincides with the physical understanding that the effects we discuss in the paper are actually caused by the physical fields in the correct direction.

A more complete description can be found in \cite{gerigk2013rf}.

\end{document}